\documentclass[manuscript]{acmart} 
\usepackage{makecell}
\usepackage{booktabs}
\usepackage{graphicx}
\usepackage{adjustbox}
\usepackage{balance}
\usepackage{marvosym}
\usepackage{subfigure}
\usepackage{multirow}
\usepackage{float}
\restylefloat{table}
\usepackage{blindtext}

\usepackage{xpatch}

\makeatletter
\xpatchcmd{\ps@firstpagestyle}{Manuscript submitted to ACM}{}{\typeout{First patch succeeded}}{\typeout{first patch failed}}
\xpatchcmd{\ps@standardpagestyle}{Manuscript submitted to ACM}{}{\typeout{Second patch succeeded}}{\typeout{Second patch failed}}    \@ACM@manuscriptfalse% Also in titlepage
\makeatother

\renewcommand\footnotetextcopyrightpermission[1]{} % removes footnote with conference info
\setcopyright{none}
\begin{document}

\title{Understanding Twitter Engagement with a Click-Through Rate-based Method}

% altri titoli
%%%% Understanding Twitter Engagement with a Click-Through Rate-based Model
%%%% Understanding Twitter Engagement with A Click-Through Rate-based Prediction Model
%%%% Understanding Twitter Engagement with a Click-Through Rate-based Model
%%%% Estimating Twitter Engagement with a Click-Through Rate-based Model
%%%% Predicting Twitter Engagement with a Click-Through Rate-based Model

\author{Andrea Fiandro}
\email{andrea.fiandro@fitec.it}
\orcid{0000-0001-8784-2574}
\affiliation{%
  \institution{FITEC srl}
  \streetaddress{Via Goito 51/A}
  \city{Grugliasco}
  \state{Italy}
  \postcode{10095}
}

\author{Jeanpierre Francois}
\email{jpr.francois@outlook.com} 
\affiliation{%
  %\institution{Politecnico di Torino}
  %\streetaddress{Corso Duca degli Abruzzi, 24}
  %\city{Turin}
  %\state{Italy}
  %\postcode{10129}
}

\author{Isabeau Oliveri}
\email{isabeau.oliveri@polito.it}
\orcid{0000-0001-5985-3583}
\affiliation{
  \institution{Politecnico di Torino}
  \streetaddress{Corso Duca degli Abruzzi, 24}
  \city{Turin}
  \state{Italy}
  \postcode{10129}
}

\author{Simone Leonardi}
\email{simone.leonardi@polito.it}
\orcid{0000-0002-8009-1082} 
\affiliation{
  \institution{Politecnico di Torino}
  \streetaddress{Corso Duca degli Abruzzi, 24}
  \city{Turin}
  \state{Italy}
  \postcode{10129}
}

\author{Matteo A. Senese}
\email{matteo.senese@linksfoundation.com}
\orcid{0000-0003-4252-7612}
\affiliation{
  \institution{LINKS Foundation}
  \streetaddress{Via Boggio, 61}
  \city{Turin}
  \state{Italy}
  \postcode{10138}
}

\author{Giorgio Crepaldi}
\email{giorgio.crepaldi@iit.it}
\orcid{0000-1234-5678-9012}
\affiliation{
  \institution{Istituto Italiano di Tecnologia}
  \streetaddress{P.O. Box 1212}
  \city{Genova}
  \state{Italy}
  \postcode{43017-6221}
}

\author{Alberto Benincasa}
\email{alberto.benincasa@linksfoundation.com}
\orcid{0000-0001-6470-2835}
\affiliation{
  \institution{LINKS Foundation}
  \streetaddress{Via Boggio, 61}
  \city{Turin}
  \state{Italy}
  \postcode{10138}
}

\author{Giuseppe Rizzo}
\email{giuseppe.rizzo@linksfoundation.com}
\orcid{0000-0003-0083-813X}
\affiliation{
  \institution{LINKS Foundation}
  \streetaddress{Via Boggio, 61}
  \city{Turin}
  \state{Italy}
  \postcode{10138}
}

\renewcommand{\shortauthors}{Fiandro et al.}

\begin{abstract}
This paper presents the POLINKS solution to the RecSys Challenge 2020\footnote{http://www.recsyschallenge.com/2020/} that ranked 6th in the final leaderboard.\footnote{https://github.com/andreafiandro/recsys2020/} 
We analyze the performance of our solution that utilizes the click-through rate value to address the challenge task, we compare it with a gradient boosting model, and we report the quality indicators utilized for computing the final leaderboard. 
\end{abstract}




\maketitle

\section{Introduction}
\label{sec:intro}
Twitter is a valuable data source for many research topics due to the richness of data it provides and the developed and recorded social interactions.
The RecSys Challenge 2020 addresses the prediction tasks of four types of user engagements on Twitter. For privacy reasons the dataset provided in the challenge is an artificial one: it is collected in one week span and consists of public engagements along with pseudo negatives randomly sampled from the public follow graph~\cite{organizersrecsys}. 
The artificial estimation of pseudo negatives along with the unbalance of the four engagement classes make the prediction task difficult for learning methods. 
We present the results of our solution implemented utilizing a method based on Click-Through Rate (CTR) on the metrics provided by the challenge and we compare them with those obtained with a gradient boosting learning model. 
We observe that our solution outperforms the learning method by a margin on the dataset provided. 

\subsection{Dataset insights}

\begin{figure}[H]
    \includegraphics[scale=0.4]{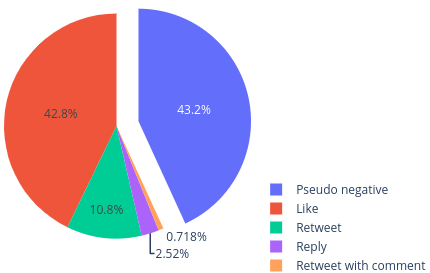}
    \caption{The chart shows the distribution of each action in the training set. Retweet with comment and Reply are very unbalanced, while pseudo negatives represent a large part of the dataset. This chart refers to the official and latest training set released.}
    \label{fig:action_distribution}
\end{figure} 

\begin{figure}[H]
    \centering
    \includegraphics[scale=0.3]{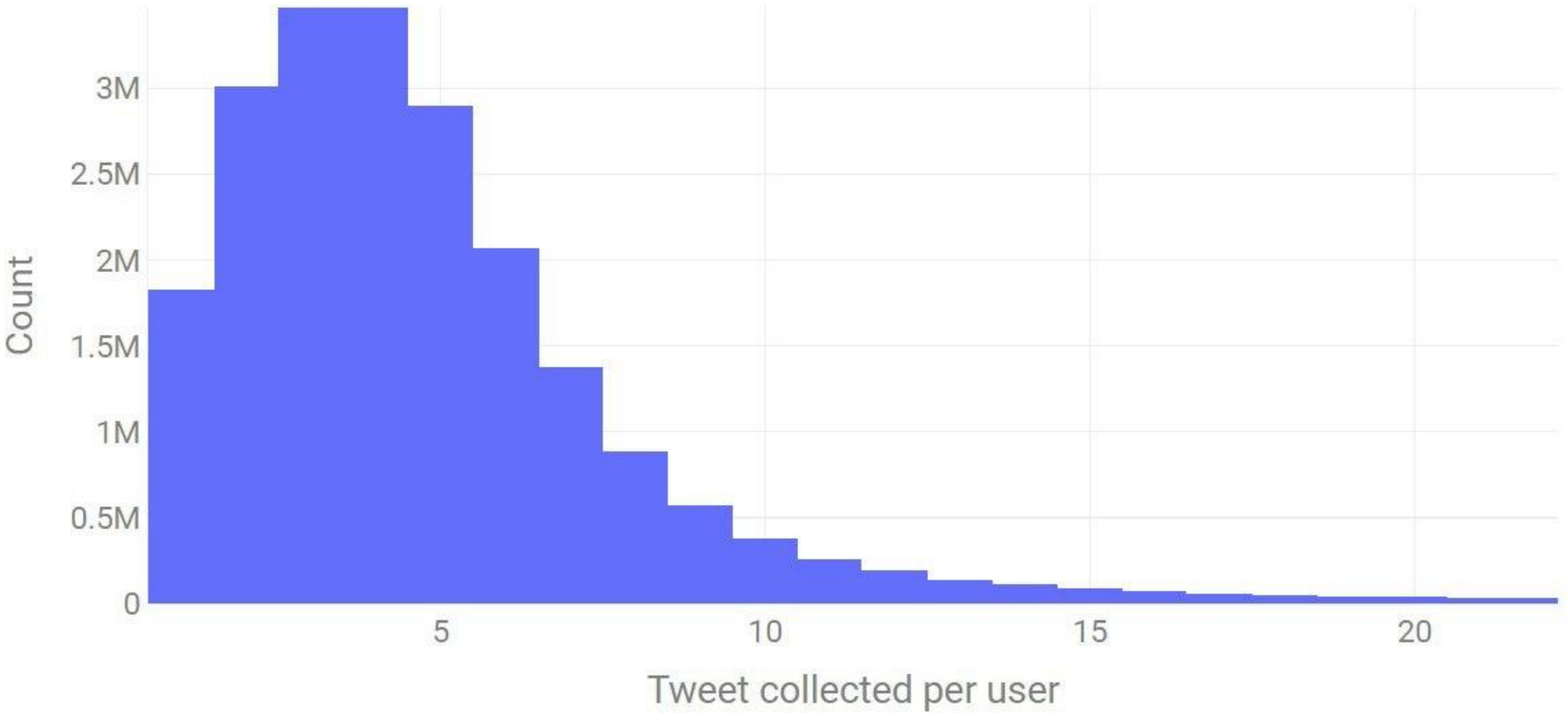}
    \caption{The histogram in figure represents the amount of tweets for which a user appears in the challenge dataset as content consumer. The horizontal axis represents how many tweets are paired with a unique user, while the vertical axis shows how many users have that number of interactions (both positive or negative).}
    \label{fig:tcpu}
\end{figure}

The dataset is a collection of public interactions on tweets along with information about their author and the user that generates the engagement. This dataset has an uneven class distribution.
As illustrated in Figure ~\ref{fig:action_distribution}, class unbalance in the training set makes the classification process difficult in both validation and test sets.
This condition is further stressed in the modality by which pseudo-negative features are obtained as described in~\cite{organizersrecsys}. In that work, authors explain the difficulties in including pseudo-negatives, samples that represent interactions with no engagement. However, the collection of this data hides the reason why a user did not interact with a tweet. In fact, a user could not interact willingly or because he did not see the tweet at all. This implies that a binary classifier is potentially misled in considering negative class candidates. 
Additionally, users' past history absence leads to the avoidance of user-based and personalized recommendation algorithms. The lack of user historical data is presented in the histogram in Figure \ref{fig:tcpu} that highlights how the majority of users interact at most with less than three tweets.

\subsection{Proposed metrics}

The organizers of the RecSys Challenge 2020 proposed two different metrics to evaluate the solutions:
\begin{description}
    \item[PRAUC] (Precision Recall Area Under the Curve)
    \item[RCE] (Relative Cross Entropy)
\end{description}
The PRAUC is useful to deal with unbalanced classes like \textit{Retweet with comment} and \textit{Reply}. These classes have numerous rows with null values. This condition means that no action is performed.
The final ranking is computed in different steps:
\begin{itemize}
    \item Averaging the PRAUC score across the four engagements
    \item Averaging the RCE score across the four engagements
    \item Compute the ranking for both metrics
    \item Sum the two obtained ranking
\end{itemize}

As we have observed in our experimental results, this way of computing the score favors solutions with a good score on the least competitive metric rankings.

\section{Our solution}
For the final submission, we propose two  solutions that we assess in terms of performance score as observed as results on the evaluation set.
In particular we submit: 
\textit{a CTR (Click-Through Rate-based) method} and
\textit{a gradient boosting model}.
This choice is supported by the following logical reasoning: the \textit{gradient boosting model} performs very well on our local test set but it has a significant worsening on the public leaderboard measured on the evaluation set. On the other hand, the CTR method behaves almost in the same way on both sets. This method used an optimized constant that is exactly the CTR value for each class. We report them in detail in the two following sections.

\subsection{Click-Through Rate-based method}
We estimate which constant value has the best outcome on both PRAUC and RCE to get a better understanding of the evaluation metrics.

The result of this investigation generates two different outcomes:
\begin{itemize}
    \item \textbf{PRAUC}: any constant produces the same effect in terms of score.
    \item \textbf{RCE}: has different outcomes depending on the engagement's type.
\end{itemize}
The best result, as pointed out by the way this metric is calculated, is given by a Click-Through Rate specific for the type of engagement.
The CTR represents the ratio of positive engagements to the number of times a tweet has been viewed.
This value is calculated, on the training set, in different steps:
\begin{itemize}
    \item Count the number of positive engagements for each class: an engagement is considered \textit{positive} if the timestamp of the interactions between a user and a tweet is not null.
    \item Count the total number of rows of the training set: this includes the four types of engagement along with pseudo-negatives.
    \item The CTR is calculated with the following equation depending on the class c:
    
    \begin{equation}
        CTR = \frac{engagement_{c}}{N_{rows}}
        \label{eq:ctr}
    \end{equation}

\end{itemize}
In Equation \ref{eq:ctr}, \textit{c} represents one of the four engagements to predict: Like, Retweet, Reply, Retweet with comment.
 The CTR numerical values are pointed out in the Table ~\ref{tab:ctr-training}.

\begin{table}[H]
\begin{tabular}{|c|c|c|c|c|}
\hline
             & \textbf{Like} & \textbf{Reply} & \textbf{Retweet} & \textbf{Retweet with comment} \\ \hline
\textbf{CTR} & 0.428         & 0.025          & 0.108            & 0.007                         \\ \hline
\end{tabular}
\caption{CTR values calculated on the training set: the class with the highest ratio is the Like that is intuitively the most frequent.}
\label{tab:ctr-training}
\end{table}

\begin{figure}
    \includegraphics[scale=0.1]{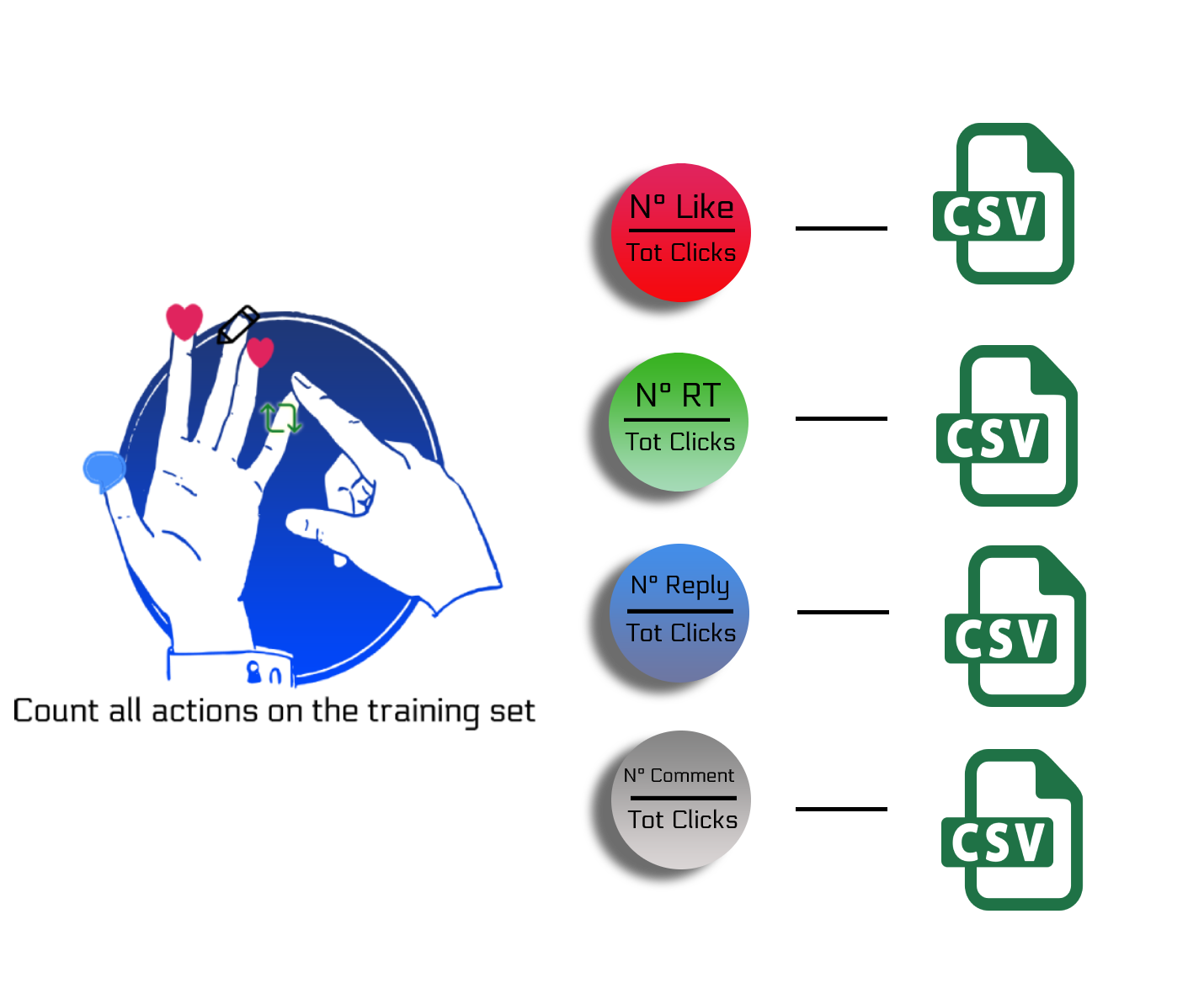}
    \caption{Overview of the model based on the CTR constant: the first step is to compute the number of actions of each type, then the constant is calculated, with a different value for each class.}
    \label{fig:constant_teaser}
\end{figure}

The optimum value is found, for both RCE and PRAUC, comparing the results of different constant value on the training set. The optimum value is the CTR in each class. The CTR scores are computed as illustrated in Figure \ref{fig:constant_teaser} where the sum of existing engagements per class (timestamp exists) over the total existing interaction per class (timestamp exists or it is a null value) represents the related CTR value.
The score is an average of the performance obtained over different chunks of the dataset.
RCE and PRAUC for the four different classes of engagement are listed in Table~\ref{tab:constant_tuning}. The second line reports the scores obtained with a random constant that is used as baseline. As shown by the other rows in Table \ref{tab:constant_tuning} when the constant values increase its absolute distance from the CTR value, RCE decreases in all the four classes.

\begin{table*}[!ht]
\begin{tabular}{@{}ccccccccc@{}}
\toprule
 & \multicolumn{4}{c}{RCE} & \multicolumn{4}{c}{PRAUC} \\ \midrule
     & Like & Reply & Retweet & \makecell{Retweet\\ with comment} & Like & Reply & Retweet & \makecell{Retweet\\ with comment} \\
    \midrule
\textbf{CTR}    & -0.01             & -0.002             & -0.003               & -0.001                            & 0.72                & 0.51                 & 0.554                  & 0.503                               \\ 
\textbf{Random} & -46.09            & -739.49            & -189.84              & -2219.446                         & 0.43                & 0.03                 & 0.109                  & 0.007                               \\ 
\textbf{0}      & -2091.49          & -642.44            & -994.003             & -483.57                           & 0.72                & 0.51                 & 0.554                  & 0.503                               \\ 
\textbf{0.1}    & -54.81            & -35.68             & -0.135               & -181.54                           & 0.72                & 0.51                 & 0.554                  & 0.503                               \\ 
\textbf{0.3}    & -5.87             & -217.64            & -30.219              & -741.73                           & 0.72                & 0.51                 & 0.554                  & 0.503                               \\
\textbf{0.5}    & -1.26449          & -481.89            & -100.908             & -1507.97                          & 0.72                & 0.51                 & 0.554                  & 0.503                               \\ 
\textbf{1}      & -2754.47          & -28153.38          & -8817.25             & -79441.8                          & 0.72                & 0.51                 & 0.554                  & 0.503                               \\
\bottomrule
\end{tabular}
\caption{Evaluation of different constants averaged across distinct portions of the training set: the CTR outperforms all the other numbers on RCE, while the PRAUC behaves in the same way with all the numbers provided.}
\label{tab:constant_tuning}
\end{table*}

\begin{figure}[ht]
\begin{subfigure}
    \centering
    \includegraphics[width=.45\linewidth]{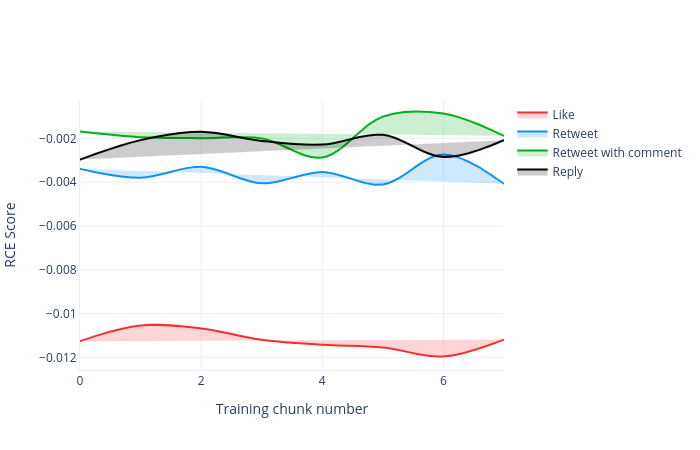}
\end{subfigure}
\begin{subfigure}
    \centering
    \includegraphics[width=.45\linewidth]{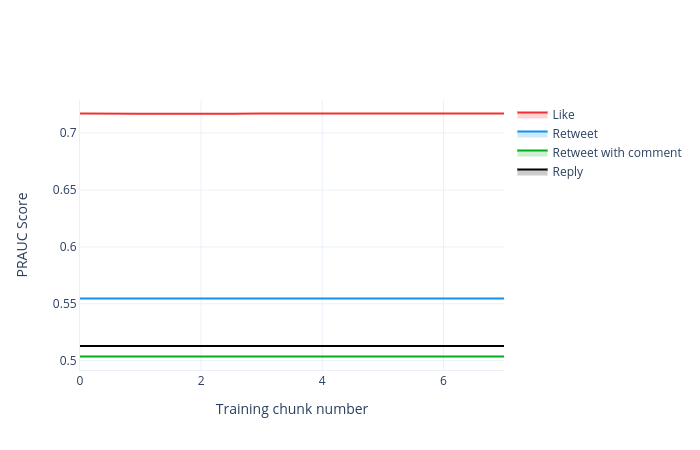}
\end{subfigure}
    \caption{The two charts describe the score obtained by the CTR constant over different time spans: the PRAUC has always the exact same value for each type of engagement and the RCE has almost the same score as well.}
    \label{fig:constant_performance}
\end{figure}

This constant value was tested on different partitions of the full training set to assess the validity of the approach.
Each partition contains 16 million rows to make them similar to the size of the challenge's validation and test set.
This number has always the same performance in terms of RCE and PRAUC throughout the different time's spans, as highlighted in Figure~\ref{fig:constant_performance}.

\subsection{Gradient boosting}

The gradient boosting approach is implemented using the XGBoost~\cite{xgboost} library and includes four different models, one for each engagement to predict.
The input of this model is enriched with 59 features that are reported in the following section. 

\subsubsection{Feature engineering}
59 additional features were derived from the dataset provided by the challenge organizers, in strict adherence with the terms and conditions of the challenge.
These additional features are grouped in different categories to facilitate their understanding:

\textbf{Dataset features (12 features)}: given directly by the dataset, they are exploited with little or no adjustments. Examples of these features are the number of hashtags, the language of the tweet and the number of followers. 

\textbf{Author features (18 features)}: this group profiles each \textit{tweet author} included in the training set. They are pre-computed features detailing the behaviour of each author during the history documented by the dataset. The most relevant features belonging to this category are:
\begin{itemize}
    \item \textit{Author engagement ratio}: $\frac{N_{eng_c}}{N_{tweet}}$ where $N_{eng_c}$ represents the number of actions of a particular type of engagement \textit{c}, where c is one among \textit{Like, Retweet, Reply, Retweet with comment}, received by a specific tweet author. Instead, the denominator $N_{tweet}$ refers to the total number of tweets published by him. In the end, there are four \textit{author engagement ratio}, one per engagement type. 
    \item \textit{Number of received engagement}: expresses the total number of interactions received by the user for each type of engagement.
\end{itemize}

\textbf{User features (18 features)}: similar to the one applied for the authors but with the involvement of the person interacting with the tweet. In this group, we find statistical features such as the \textit{engagement ratio} and the \textit{number of actions for each type}, calculated from the user point of view. 

\textbf{Languages spoken (1 feature)}: the main intuition behind this feature is that understanding the language of a tweet plays a key role for a user in having possible interactions.
This approach includes the pre-computation of a file containing, for each user id, the number of times that a specific user has interacted with a tweet written in a specific language.
The goal of this computation is to identify, for each user $U_{id}$, a list of languages \textit{spoken} by that specific user.
In more formal terms:
\begin{equation}
    f(U_{id}, Lang_{id}) = n_{LT}
\end{equation}
where $Lang_{id}$ is the id of a language and $n_{LT}$ is the number of tweets written in that language engaged by the user.

\textbf{Previous actions (4 features)}: another pre-computation is performed to reconstruct the history of previous interactions. This set of features are formalized with the following function:
\begin{equation}
f(U_{id}, A_{id}, c) = n_{PA}
\end{equation}

where: 
\begin{description}
    \item[$U_{id}$] is the user id
    \item[$A_{id}$] is the author id
    \item[$c$] is the class representing the engagement type
    \item[$n_{PA}$] is the number of previous actions for the triplet $(U_{id}, A_{id}, c)$
\end{description}

\textbf{Word search (6 features)}
This class of features is the only one referring to the text of the tweet. We extract some meaningful words from the text tokens and generate a boolean variable to identify when that specific word is included in the text of the tweet.
The words used are related to the \textit{call to action}, a situation when the tweet author invites his followers to do a specific action with respect to the tweet.
The considered words are \textit{share}, \textit{retweet}, \textit{reply}, \textit{comment}.

\section{Submission}

\subsection{Click-Through Rate-based}
This submission was performed using the value of the CTR calculated on the whole training set. The intuition behind this approach was that, if the distribution of positive actions with respect to the negative ones does not change too much, we can achieve a score that outperforms several proposed models including the \textit{gradient boosting model}. 

\subsection{Gradient Boosting}
The final model includes almost sixty different features.
The early stopping feature of the XGBoost library was used to avoid overfitting. After each training round the model is evaluated on a validation set using the RCE metric and if there is no improvement with respect to the last N round the training is stopped.

The four models were trained on the final release of the dataset with the parameters in Table \ref{tab:xgboostparam}.
\begin{table}[H] 
\begin{tabular}{|l|l|l|l|l|l|}
\hline
\textbf{Parameter}      & \textbf{Value} & \textbf{Parameter}      & \textbf{Value} & \textbf{Parameter}      & \textbf{Value}  \\ \hline
eta                     & 0.09            &
tree\_method            & gpu\_hist       &
sampling\_method        & gradient\_based \\ \hline
subsample               & 0.2             &
objective               & binary:logistic &
max\_depth              & 5               \\ \hline
max\_delta\_step        & 5               &
epochs                  & 200             &
early\_stopping\_rounds & 10              \\ \hline
\end{tabular}
\caption{XGBoost parameters used by the four different models during the training phase.}
\label{tab:xgboostparam}
\end{table}

\subsection{Results}
The optimized constant based model achieves better results when compared to those obtained with gradient boosting. The results are reported in Table \ref{tab:leaderboard}. The reason why a  constant-based method performs better than a gradient boosted one is due to the way score and ranking are computed in the RecSys Challenge 2020. As presented by the winning team of the challenge, a very computational heavy, more complex and deeply parallel boosting model is able to explore each row in the dataset characteristics, while our fine tuned XGBoost on a subset of the dataset does not. In this way, an optimized constant is able to generalize better the classification procedure avoiding overfitting with respect to the gradient boosted one as described both in Figure \ref{fig:constant_performance} and in Table \ref{tab:leaderboard}.

\begin{table*}[!ht]
\begin{tabular}{@{}cccccccccc@{}}
\toprule
& & \multicolumn{2}{c}{Retweet} & \multicolumn{2}{c}{Reply} & \multicolumn{2}{c}{Like} & \multicolumn{2}{c}{RwC} \\ \midrule
Model & Dataset & PRAUC & RCE & PRAUC & RCE &  PRAUC & RCE & PRAUC & RCE \\ \midrule
\multirow{3}{*}{CTR-based} & \makecell{Final Leaderboard\\ (Test set)} &  0.5516 & -0.03 & 0.5135	& -0.05 & 0.7131 & 0	& 0.5037 & 0\\ 
 & \makecell{Public Leaderboard\\ (Validation set)} & 0.5516 & -0.0315 & 0.5135 & -0.0476 & 0.7133  & -0.0008  & 0.5037  & -0.0045  \\  
 &  Local test set & 0.72 & -0.01 & 0.51 & -0.002 & 0.554 & -0.003 & 0.503 & -0.001 \\  \midrule
\multirow{2}{*}{XGBoost} & \makecell{Public Leaderboard\\ (Validation set)} &  0.41 & 6.68 & 0.10 & -3.23 & 0.66 & -32.01 & 0.04 & -1.67\\ 
 & Local test set & 0.710 & 46.85 & 0.626 & 35.24 & 0.830 & 32.56 & 0.586 & -8.916 \\
\bottomrule
\end{tabular}
\caption{Results of the two described models in different contexts: while the CTR constant maintains almost the same score, the XGBoost model loses efficiency when the time difference with respect to the training set period increases.}
\label{tab:leaderboard}
\end{table*}

\section{Final Consideration} 
The CTR-based method addresses the issues related to the unbalanced classes and pseudo negatives as described in Section \ref{sec:intro}. However, the literature on recommender systems \cite{quadrana,dacremadeep,recissues,netflixrec,reccase} highlights some issues that are intrinsically related to the problems addressed in this challenge and, therefore, observable in the dataset provided.

\textit{Short term trends}: those trends that tend to change or disappear quickly due to the rapid evolution of individual and community preferences.

\textit{Cold start problem}: when new users enter the system, the preferences of the new users could not be predicted.

\textit{Gray sheep}: problematical users that cannot be traced or predictably aligned to any trend, so the suggestions related to current trends are not an effective solution.

\textit{Real-time analysis}: real-time data have to be collected to perform analysis for unexpected events (e.g., earthquakes, pandemics) and a continuous update of the model with real-time information.

\textit{Context-Awareness, Privacy, and Sparsity}: users' short and long term history and unnoticeable context-related information may not be retrieved~\cite{recissues}. Despite the huge size of data typically collected, most users are occasional or not inclined to interact, both for privacy issues and unwanted exposure of information. These aspects lead to a sparse characterization matrix, thus resulting in less accurate recommendations. 

\textit{Baseline metrics}: the available evaluation metrics are for general-purpose recommender systems and are not always applicable in different domains, especially in evaluating the context-aware ones. Metrics used in common machine learning approaches do not always lead to well suited recommendations~\cite{dacremadeep}.  

The constant preserves its performance among training, validation and test set, while the XGBoost model gets worse. A model could be inefficient if it is not able to capture time and event independent features. This, along with the above issues, could be a probable cause of the considerable variation of the challenge leaderboard, from the validation to the final test phase. In fact, as reported\footnote{https://recsys-twitter.com}, the entire dataset was produced in two different weeks. Thanks to our observations over the result of the validation set, we conclude that an optimized constant performs better. This intuition turns out to be successful in the test phase, indeed, the POLINKS solution ranked sixth at the end of the RecSys Challenge 2020.

\begin{acks}
This research was supported by FITEC S.r.l., LINKS Foundation, Politecnico di Torino and Istituto Italiano di Tecnologia. Computational resources were provided by HPC@POLITO\footnote{http://www.hpc.polito.it}, a project of Academic Computing within the Department of Control and Computer Engineering at the Politecnico di Torino. 
\end{acks}

\end{document}